\documentclass[aps,twocolumn,showpacs,superscriptaddress]{revtex4-1}
\usepackage{graphicx,color}
\usepackage{amsmath}
\usepackage{amssymb}
\usepackage{amsfonts}
\usepackage{bm}

\newcommand{\ot}{{\,\otimes\,}}
\newcommand{{\Cd}}{{\mathbb{C}^d}}

\def\oper{{\mathchoice{\rm 1\mskip-4mu l}{\rm 1\mskip-4mu l}%
{\rm 1\mskip-4.5mu l}{\rm 1\mskip-5mu l}}}
\def\<{\langle}
\def\>{\rangle}
\newtheorem{Theorem}{Theorem}
\newtheorem{Lemma}{Lemma}

\newtheorem{Example}{Example}

\begin{document}

\title{On the degree of non-Markovianity of quantum evolution}
\author{Dariusz Chru\'sci\'nski$^1$ and Sabrina Maniscalco$^{2}$\\
$^1$Institute of Physics, Nicolaus Copernicus University,
Grudzi\c{a}dzka 5/7, 87--100 Toru\'n, Poland\\
$^2$SUPA, EPS/Physics, Heriot-Watt University, Edinburgh, EH14 4AS, United Kingdom }

%

\begin{abstract}

We propose a new characterization of non-Markovian quantum evolution based on the concept of non-Markovianity degree. It provides an analog of a Schmidt number in the entanglement theory and reveals the formal analogy between quantum evolution and the entanglement theory: Markovian evolution corresponds to a separable state and non-Markovian one is further characterized by its degree.  It enables one to introduce a non-Markovinity witness -- an analog of an entanglement witness -- and a family of measures -- an analog of Schmidt coefficients -- and finally to characterize maximally non-Markovian evolution being an analog of maximally entangled state. 
Our approach allows to classify the non-Markovianity measures introduced so far in a unified rigorous mathematical framework.

\end{abstract}



\maketitle

{\em Introduction} --- Open quantum systems and their dynamical features are
attracting increasing attention nowadays. They are of paramount
importance in the study of the interaction between a quantum
system and its environment, causing dissipation, decay, and
decoherence \cite{Breuer,Weiss,Alicki}. On the other hand, the robustness of
quantum coherence and entanglement against the detrimental
effects of the environment is one of the major focuses in
quantum-enhanced applications, as both entanglement and
quantum coherence are basic resources in modern quantum
technologies, such as quantum communication, cryptography,
and computation \cite{QIT}. Recently, much effort was devoted to {the} description, analysis and classification of non-Markovian quantum evolution (see e.g. \cite{R1}--\cite{Pater} and the collection of papers in  \cite{RECENT}). In particular various concepts of non-Markovianity were introduced and several so called non-Markovianity measures were proposed. The main approaches to the problem of (non)Markovian evolution are based on divisibility \cite{Wolf,RHP,Hou}, distinguishability of states  \cite{BLP},
quantum entanglement \cite{RHP}, quantum Fisher information flow \cite{Fisher}, fidelity \cite{fidelity}, mutual information \cite{Luo1,Luo2}, channel capacity \cite{Bogna}, and  geometry of the set of accessible states \cite{Pater}.

In this Letter we accept the  definition based on divisibility \cite{Wolf}: the quantum evolution is Markovian if the corresponding dynamical map $\Lambda_t$ is CP-divisible, that is,
\begin{equation}\label{CP}
  \Lambda_t = V_{t,s} \Lambda_s \ ,
\end{equation}
and $V_{t,s}$ provides a  family of legitimate (completely positive and trace-preserving) propagators for all $t\geq s\geq 0$.
The essential property of $V_{t,s}$ is the following composition law
 $   V_{t,s} \, V_{s,u} = V_{t,u}$,
for all $t\geq s\geq u$. It provide{s} a natural generalization of a semigroup law $e^{tL} e^{sL} = e^{(t+s)L}$. Interestingly, the very property of CP-divisibility is fully characterized in terms of the time-local generator $L_t$:  if $\Lambda_t$ satisfies time-local master equation $\dot{\Lambda}_t = L_t \Lambda_t$, then $\Lambda_t$ is CP-divisible iff $L_t$ has the standard Lindblad form for all $t \geq 0$, i.e.
$$  L_t\rho = -i[H(t),\rho] + \sum_\alpha \left( V_\alpha(t) \rho V_\alpha^\dagger(t) - \frac 12 \{ V_\alpha^\dagger(t) V_\alpha(t),\rho\} \right), $$
with time-dependent Lindblad (noise) operators $V_\alpha(t)$ and time-dependent effective system Hamiltonian $H(t)$ \cite{GKS,Lindblad,Alicki}. A very appealing concept of Markovianity was proposed  by Breuer, Lane and Piilo (BLP)  \cite{BLP}: $\Lambda_t$ is Markovian if
\begin{equation}\label{BLP}
\sigma(\rho_1,\rho_2;t) = \frac{d}{dt}||\Lambda_t(\rho_1-\rho_2)||_1 \leq 0 \ ,
\end{equation}
for all pairs of initial states $\rho_1$ and $\rho_2$. BLP call $\sigma(\rho_1,\rho_2;t)$ an information flow and interpret $\sigma(\rho_1,\rho_2;t)> 0$ as a backflow of information from the environment to the system which clearly indicates the non-Markovian character of the evolution. As usual $||X||_1$ denotes the trace norm of $X$, i.e.  $||X||_1 = {\rm Tr}\sqrt{XX^\dagger}$. It turns out that CP-divisibility implies (\ref{BLP}) but the converse needs not be true \cite{versus1,versus2,versus3}.

In this Letter we propose {a} more refine{d} approach to non-Markovian evolution. We reveal the formal analogy with the entanglement theory: Markovian evolution corresponds to separable state and non-Markovian evolution is characterized by {a} positive integer ---  {the} non-Markovianity degree --- corresponding to the Schmidt number of an entangled state. The notion of non-Markovianity degree enables one to introduce a family of measures and finally to characterize maximally non-Markovian evolution being an analog of maximally entangled state.

{\em Schmidt number and $k$-positive maps} --- Let us recall that a state of {a} composite quantum system may be uniquely characterized by its Schmidt number \cite{HHHH,Pawel}: for any normalized vector $\psi \in \mathcal{H} \ot \mathcal{H}$ let ${\rm SR}(\psi)$ denote the Schmidt rank of $\psi$, i.e. a number of non-vanishing Schmidt coefficients in the decomposition $\psi = \sum_k s_k e_k \ot f_k$, with $s_k > 0$ and $\sum_k s_k^2=1$. Now, for any density operator  $\rho$ one defines its  Schmidt number by
\begin{equation}\label{}
  {\rm SN}(\rho) = \min_{p_k,\psi_k}\, \{ \max_k \, {\rm SR}(\psi_k) \}\ ,
\end{equation}
where the minimum is performed over all decompositions $\rho = \sum_k p_k |\psi_k\>\<\psi_k|$ with $p_k >0$ and $\sum_k p_k=1$. Let $S_k= \{ \, \rho\, |\, {\rm SN}(\rho) \leq k\, \}$. One has
\begin{equation}\label{SSS}
  S_1 \subset S_2 \subset \ldots \subset S_n \ ,
\end{equation}
where $S_1$ denotes a set of separable states and $S_n$ denotes a set of all states in $\mathcal{H} \ot \mathcal{H}$. Note that a maximally entangled state $\psi$ satisfies $\lambda_1 = \ldots = \lambda_n$ and the corresponding projector $|\psi\>\<\psi|$ defines an element of $S_n$. The Schmidt number does not increase under local operation, i.e. ${\rm SN}([\mathcal{E}_1 \ot \mathcal{E}_2]\rho) \leq  {\rm SN}(\rho)$, where $\mathcal{E}_1$ and $\mathcal{E}_2$ are arbitrary quantum channels. Moreover, if $\Phi$ is a $k$-positive map, i.e. $\oper_k \ot \Phi$ is positive, then for any $\rho \in S_k$ one has $[\oper_k \ot \Phi](\rho) \geq 0$ ($\oper_k$ denotes an identity map acting in $M_k$ --- the space of $k\times k$ complex matrices).  This simple property establishes a duality between  $k$-positive maps and quantum bipartite states with the Schmidt number bounded by $k$.

{\em Non-Markovianity degree} --- The notion of $k$-positive maps enables one to provide a natural generalization of CP-divisibility: we call a dynamical map $\Lambda_t$ $k$-divisible iff $V_{t,s}$ is $k$-positive for all $t\geq s\geq 0$. Hence, $n$-divisible maps are CP-divisible and 1-divisible are simply P-divisible, i.e. $V_{t,s}$ is positive. Now, we introduce a degree of non-Markovianity which is an analog of a Schmidt number: a dynamical map $\Lambda_t$ has a non-Markovianity degree ${\rm NMD}[\Lambda_t]=k$ iff $\Lambda_t$ is $(n-k)$ but not $(n+1-k)$--divisible.
It is clear that $\Lambda_t$ is Markovian iff ${\rm NMD}[\Lambda_t]=0$ and essentially non-Markovian iff ${\rm NMD}[\Lambda_t]=n$. Denoting by
 $ \mathcal{N}_k = \{ \, \Lambda_t\ | \ {\rm NMD}[\Lambda_t] \leq k \, \}$,
one has a natural chain of inclusions
\begin{equation}\label{NNN}
  \mathcal{N}_0 \subset \mathcal{N}_1 \subset \ldots \subset \mathcal{N}_{n-1} \subset \mathcal{N}_n\ ,
\end{equation}
where $\mathcal{N}_0$ denotes Markovian maps and $\mathcal{N}_n$ all dynamical maps. The characterization of $k$-divisible maps is provided by the following

\begin{Theorem}
If $\Lambda_t$ is $k$-divisible, then
\begin{equation}\label{contr-k}
     \frac {d}{dt}\,  || [\oper_k \ot \Lambda_t](X) ||_1 \leq 0\ ,
\end{equation}
for all operators  $X \in M_k \ot \mathcal{B}(\mathcal{H})$.
\end{Theorem}
For the proof see Supplementary material. In particular all $k$-divisible maps ($k=1,\ldots,n$) satisfy
\begin{equation}\label{contr-1}
     \frac {d}{dt}\,  ||  \Lambda_t(X) ||_1 \leq 0\ ,
\end{equation}
for all $X \in \mathcal{B}(\mathcal{H})$. Note, that BLP condition (\ref{BLP}) is a special case of (\ref{contr-1}) with $X$ being traceless Hermitian operator. It is, therefore, clear that BLP condition is weaker than all conditions in the hierarchy (\ref{contr-k}) and it is satisfied for all $k$-divisible maps not necessarily CP-divisible. According to our definition of Markovianity  (Markovianity = CP-divisibility) $k$-divisible maps which are not CP-divisible are clearly non-Markovian. However, such  non-Markovian evolution always satisfy (\ref{contr-1}). We propose to call such dynamical maps  {\em weakly non-Markovian}. Dynamical map which is even not P-divisible will be called {\em essentially non-Markovian}. Hence, $\Lambda_t$ is weakly non-Markovian iff $\Lambda_t \in \mathcal{N}_{n-1} - \mathcal{N}_0$ and it is essentially non-Markovian iff  $\Lambda_t \in \mathcal{N}_{n} - \mathcal{N}_{n-1}$. Using the notion of degree of non-Markovianity $\Lambda_t$ is weakly non-Markovian iff $0< {\rm NMD}[\Lambda_t] \leq n-1$ and it is essentially non-Markovian iff ${\rm NMD}[\Lambda_t] =n$. Note that maps which violate BLP condition are always essentially non-Markovian. Similarly, if $\Lambda_t$ is at least 2-divisible, then the relative entropy satisfies the following monotonicity property \cite{Petz}
\begin{equation}\label{RE}
     \frac {d}{dt}\,  S( \Lambda_t(\rho_1) || \Lambda_t(\rho_2)) \leq 0\ ,
\end{equation}
for any pair $\rho_1$ and $\rho_2$.  The violation of (\ref{RE})  means that $\Lambda_t$ is at most P-divisible or essentially non-Markovian.

{\em Non-Markovianity witness} --- Actually, if  $\Lambda_t$ is invertible, then it is $k$-divisible if and only if (\ref{contr-k}) holds.  Clearly, a generic map is invertible  (all its eigenvalues are different from zero) and hence this result is true for a generic dynamical map (a notable exception is Jaynes-Cummings model on resonance \cite{Breuer,JC}).
Hence, if (\ref{contr-k}) is violated for some $t > 0$, then $\Lambda_t$ is not $k$-divisible or equivalently ${\rm NMD}[\Lambda_t] > n-k$. It is, therefore, natural to call such $X$ a non-Markovianity witness in analogy to the well known concept of an entanglement witness. Recall, that a Hermitian operator $W$ living in $\mathcal{H} \ot \mathcal{H}$ is an entanglement witness \cite{HHHH} iff $i)$ $\<\Psi|W|\Psi\> \geq 0$ for all product vectors $\Psi=\psi\ot \phi$ and $ii)$ $W$ is not a positive operator, i.e. it possesses at least one negative eigenvalue. Similarly, $W$ is a $k$-Schmidt witness \cite{Sanpera} if $\<\Psi|W|\Psi\> \geq 0$ for all vectors $\Psi=\psi_1\ot \phi_1 + \ldots + \psi_k \ot \phi_k$, that is, if ${\rm Tr}(\rho W) <0$, then $\rho$ is entangled and moreover ${\rm SN}(\rho) > k$.  Note, that if $X \geq 0$, then (\ref{contr-k}) is always satisfied due to the fact that
$|| [\oper_k \ot \Lambda_t](X) ||_1 = ||X||_1$. Hence, similarly as $W$, a non-Markovianity witness $X$ has to possess a negative eigenvalue.

{\em Non-Markovianity measures} --- The above construction allows to define a series of natural measures measuring departure from $k$-divisibility:
 \begin{equation}\label{}
   {\cal M}_k[\Lambda_t] = \sup_{X} \frac{N_k^+[X]}{|N_k^-[X]|} \ ,
 \end{equation}
where
\begin{eqnarray*}\label{}
  N^+_k[X] &=& \int_{\lambda_k(X;t) >0} \lambda_k(X;t) dt\  , 
\end{eqnarray*}
and similarly for $ N^+_k[X]$ (where now one integrates over time intervals such that $\lambda_k(X;t) < 0$), and
\begin{equation}\label{}
   \lambda_k(X;t) = \frac {d}{dt}\,  || [\oper_k \ot \Lambda_t](X) ||_1\ .
\end{equation}
The supremum is taken over all Hermitian $X \in M_k \ot \mathcal{B}(\mathcal{H})$.
Note that
\begin{eqnarray*}
&& \int_0^\infty \frac {d}{dt}\,  || [\oper_k \ot \Lambda_t](X) ||_1 \, dt  \\ && = ||[\oper_k \ot \Lambda_\infty](X) ||_1 - ||X||_1 \leq 0\ ,
\end{eqnarray*}
and hence $|N_-[\Lambda_t]| \geq N_+[\Lambda_t|$ which proves that  $\mathcal{M}_k[\Lambda_t] \in [0,1]$. Clearly, if $l >k$, then  $\mathcal{M}_l[\Lambda_t] \geq \mathcal{M}_k[\Lambda_t]$ and hence
 \begin{equation*}
0 \leq  \mathcal{M}_1[\Lambda_t] \leq \ldots \leq \mathcal{M}_n[\Lambda_t] \leq 1\ ,
 \end{equation*}
which provides an analog of a similar relation among the Schmidt coefficients $s_1\geq \ldots \geq s_n$. Now, following the analogy with an entanglement theory, we may call $\Lambda_t$ maximally non-Markovian iff $\mathcal{M}_1[\Lambda_t]=1$ which immediately implies
\begin{equation}\label{}
 \mathcal{M}_1[\Lambda_t] = \ldots = \mathcal{M}_n[\Lambda_t] =1\ ,
\end{equation}
in a perfect analogy with maximally entangled state corresponding to $s_1 = \ldots = s_n$.

{\em Examples} --- Let us illustrate the above introduced notions by a few simple examples.

\begin{Example}
Consider pure decoherence of a qubit system described by the following local generator
\begin{equation}\label{pure}
    L_t(\rho) = \frac 12 \gamma(t) (\sigma_z \rho \sigma_z - \rho) \ ,
\end{equation}
The corresponding evolution of the density matrix reads
\begin{equation}\label{}
    \rho_t = \left( \begin{array}{cc} \rho_{11} & \rho_{12} e^{-\Gamma(t)} \\ \rho_{12} e^{-\Gamma(t)} & \rho_{22} \end{array} \right) \ ,
\end{equation}
where $\Gamma(t) = \int_0^t \gamma(\tau) d\tau$. The evolution is completely positive iff $\Gamma(t) \geq 0$ and it is $k$-divisible ($k=1,2$)  iff $\gamma(t) \geq 0$. Taking $X = \sigma_x $ one finds $||\Lambda_t(X)||_1 = 2e^{-\Gamma(t)}$. Observe that
\begin{equation}\label{}
  |N_-[\Lambda_t]| = N_+[\Lambda_t] + e^{-\Gamma(\infty)} - 1 \ ,
\end{equation}
and hence if $\Gamma(\infty)=0$ the evolution is maximally non-Markovian. Note, that $\Gamma(\infty)=0$ implies that $\rho_t \rightarrow \rho$, that is, asymptotically one always recovers an initial state -- perfect recoherence. Actually, this example may be immediately generalized as follows: let $L$ be a Lindblad generator and consider a time-dependent generator defined by $L_t = \gamma(t) L$. Now, $L_t$ gives rise to a legitimate quantum dynamical map iff $\Gamma(t) \geq 0$ and it is $k$-divisible ($k=1,2,\ldots,n$) iff $\gamma(t)\geq 0$. The corresponding dynamics is maximally non-Markovian if $\Gamma(\infty)=0$.
\end{Example}

\begin{Example} Consider the qubit dynamics governed by the time-dependent generator
\begin{equation}\label{III}
    L_t(\rho) = \frac 12 \sum_{k=1}^3 \gamma_k(t) (\sigma_k \rho \sigma_k - \rho) \ .
\end{equation}
It is clear that (\ref{III}) provides simple generalization of (\ref{pure}) by introducing two additional decoherence channels.
The corresponding dynamical map reads
\begin{equation}\label{rud}
    \Lambda_t(\rho) =  \sum_{\alpha=0}^3 p_\alpha(t) \sigma_\alpha \rho \sigma_\alpha  \ ,
\end{equation}
where $\sigma_0 = \mathbb{I}$, and the probability distribution $p_\alpha(t)$ is defined as follows
\begin{eqnarray*}
p_0(t) &=&\frac{1}{4}\, [1+ \lambda_3(t) + \lambda_2(t) + \lambda_1(t)] \ , \\
p_1(t) &=&\frac{1}{4}\, [1- \lambda_3(t) - \lambda_2(t) + \lambda_1(t)] \ ,\\
p_2(t) &=&\frac{1}{4}\, [1- \lambda_3(t) + \lambda_2(t) - \lambda_1(t)] \ ,\\
p_3(t) &=&\frac{1}{4}\, [1+ \lambda_3(t) - \lambda_2(t) - \lambda_1(t)] \ ,
\end{eqnarray*}
where $\lambda_1(t) = e^{-[\Gamma_2(t) + \Gamma_3(t)]}$ and similarly for $\lambda_2(t)$ and $\lambda_3(t)$.  Finally, $\Gamma_k(t) = \int_0^t \gamma_k(\tau) d\tau$. Again, the map $\Lambda_t$ is completely positive iff $\Gamma_k(t) \geq 0$ for $k=1,2,3$. Interestingly, in this example there is an essential difference between CP-divisibility (= Markovianity) and only P-divisibility: CP-divisibility is equivalent to
\begin{equation}\label{123}
  \gamma_1(t) \geq 0 \ , \ \  \gamma_2(t) \geq 0 \ , \ \  \gamma_3(t) \geq 0 \ ,
\end{equation}
whereas P-divisibility is equivalent to much weaker conditions \cite{PLA}
\begin{eqnarray}\label{2gamma}
    \gamma_1(t) + \gamma_2(t) &\geq&  0 \ , \nonumber \\
    \gamma_1(t) + \gamma_3(t) &\geq& 0 \ , \\
    \gamma_2(t) + \gamma_3(t) &\geq& 0 \ , \nonumber
\end{eqnarray}
for all $t\geq 0$. Actually, the BLP condition reproduces (\ref{2gamma}). Now, violation of at least one inequality from (\ref{2gamma}) implies essential non-Markovianity. Suppose for example that $\gamma_2(t) + \gamma_3(t) \ngeq 0$. Assuming  that $\Gamma_2(\infty)=\Gamma_3(\infty)=0$ one finds that $ \mathcal{M}_1[\Lambda_t] =1$, that is, $\Lambda_t$ is maximally non-Markovian. Interestingly, if there are at most two decoherence channels, then there is no difference between CP- and P-divisibility. Note, that random unitary dynamics (\ref{rud}) is unital, i.e. $\Lambda_t(\mathbb{I}) = \mathbb{I}$ and hence during the evolution the entropy never decreases $S(\Lambda_t(\rho)) \geq S(\rho)$ for any initial qubit state $\rho$. One easily shows that
P-divisibility is  equivalent to
\begin{equation}\label{S1}
  \frac{d}{dt} S(\Lambda_t(\rho)) \geq 0 \ ,
\end{equation}
for any qubit state $\rho$. Hence, for any weakly non-Markovian random unitary dynamics the von Neumann entropy monotonically increases. Violation of (\ref{S1}) proves that $\Lambda_t$ is essentially non-Markovian.
\end{Example}

\begin{Example}
Consider a qubit dynamics governed by the following local generator
\begin{equation}\label{}
  L_t =  \gamma_+(t) L_+ + \gamma_-(t) L_-\ ,
\end{equation}
where
\begin{eqnarray*}
  L_+(\rho) &=& \frac 12 ([\sigma_+,\rho \sigma_-] +  [\sigma_+\rho, \sigma_-]) \ ,\\
  L_-(\rho) &=&  \frac 12 ( [\sigma_-,\rho \sigma_+] +  [\sigma_-\rho, \sigma_+]) \ ,
\end{eqnarray*}
{with} $\sigma_+ = |2\>\<1|$ and    $\sigma_- = |1\>\<2|$. $L_+$ generates pumping from the ground state $|1\>$ to an excited state $|2\>$ and $L_-$ generates a decay from $|2\>$ to $|1\>$. One shows that $L_t$ generates legitimate dynamical map iff
\begin{equation}\label{gg}
  0\leq \int_0^t \gamma_\pm(s) e^{\Gamma(s)}ds \leq e^{\Gamma(t)} - 1\ ,
\end{equation}
where $\Gamma(t) = \int_0^t [ \gamma_-(\tau) + \gamma_+(\tau)]d\tau$. In particular it follows from (\ref{gg}) that $\Gamma(t) \geq 0$.
Now, $\Lambda_t$ is CP-divisible iff
\begin{equation}\label{-+1}
  \gamma_-(t)\geq 0 \ , \ \ \   \gamma_+(t) \geq 0\ ,
\end{equation}
and it is P-divisible iff
\begin{equation}\label{-+2}
  \gamma_-(t) + \gamma_+(t) \geq 0\ .
\end{equation}
Note, that (\ref{-+1}) implies (\ref{gg}). However, it is not true for (\ref{-+2}), i.e. P-divisibility requires both (\ref{gg}) -- it {guarantees} that $\Lambda_t$ is completely positive -- and (\ref{-+2}).
\end{Example}

{\em Bloch equations and P-divisibility} --- The above examples illustrating qubit dynamics may be easily rewritten in terms of the Bloch vector $x_k(t) = {\rm Tr}[\sigma_k \Lambda_t(\rho)]$. Example 2 gives rise to
\begin{equation}\label{}
  \frac{d}{dt} x_k(t) = - \frac{1}{T_k(t)} x_k(t)\ , \ \ k=1,2,3\ ,
\end{equation}
where $T_1(t) = [\gamma_2(t) + \gamma_3(t)]^{-1}$, and similarly for $T_2(t)$ and $T_3(t)$.  Quantities $T_k(t)$ correspond to local relaxation times. It is therefore clear that P-divisibility is equivalent to $T_k(t) \geq 0$ for $k=1,2,3$. Hence, CP-divisibility requires that all local decoherence rates satisfy $\gamma_k(t) \geq 0$, whereas P-divisibility requires only $T_k(t) \geq 0$. Note, that CP-divisibility is equivalent to P-divisibility plus three extra conditions
\begin{equation*}\label{}
  \frac{1}{T_1} +   \frac{1}{T_2} \geq  \frac{1}{T_3}\ ,\ \ \frac{1}{T_1} +   \frac{1}{T_3} \geq  \frac{1}{T_2}\ ,\ \ \frac{1}{T_2} +   \frac{1}{T_3} \geq  \frac{1}{T_1}\ .
\end{equation*}
 Finally, let us observe that the initial volume of the
Bloch ball shrinks during the evolution according to
$$V(t) = e^{-[\Gamma_1(t) + \Gamma_2(t) + \Gamma_3(t)]} V(0)\ , $$
where $V(t)$ denotes a volume of the set of accessible states at time $t$. Authors of \cite{Pater} characterized non-Markovian evolution as a departure from $\frac{d}{dt} V(t) \leq 0$. One has   $\frac{d}{dt} V(t) = - [\gamma_1(t) + \gamma_2(t) + \gamma_3(t)] V(t)$ and hence $\frac{d}{dt} V(t) \leq 0$ iff
\begin{equation}\label{3gamma}
  \gamma_1(t) + \gamma_2(t) + \gamma_3(t) \geq 0\ .
\end{equation}
This condition is much weaker than (\ref{2gamma}). To violate (\ref{3gamma}) the evolution has to be essentially non-Markovian (i.e. $\Lambda_t$ can not be even P-divisible).

A similar conclusion may be { drawn} from Example 3: the corresponding Bloch equations read
\begin{eqnarray}\label{}
  \frac{d}{dt}\, x_1(t) &=& - \frac{1}{T_\perp(t)}\, x_1(t)\ , \nonumber \\
  \frac{d}{dt}\, x_2(t) &=& - \frac{1}{T_\perp(t)}\, x_2(t)\ , \\
  \frac{d}{dt}\, x_3(t) &=& - \frac{1}{T_{||}(t)}\,  x_3(t) + \Delta(t)\ ,\nonumber
\end{eqnarray}
where $\Delta(t) = [\gamma_+(t) - \gamma_-(t)]$, and $T_\perp(t) = 2/[\gamma_-(t) + \gamma_+(t)]$ and $T_{||}(t) = T_\perp(t)/2$ are transverse and longitudinal local relaxation times, respectively. Again, P-divisibility is equivalent to $T_\perp,  T_{||}(t) \geq 0$ provided that the Bloch vector stays within a Bloch ball.

{\em Conclusions} --- In this Letter we provided further characterization of non-Markovian evolution in terms of non-Markovianity degree. This simple concept, being an analog of the Schmidt number in the entanglement theory, enables one to compare quantum evolutions and finally to define a maximally non-Markovian evolution being an analog of a maximally entangled state.


{\em Acknowledgements} --- Our research was partially completed while the authors were visiting the Institute for Mathematical Sciences, National University of Singapore { in the framework of the programme Mathematical Horizons for Quantum Physics 2} in 2013.
D.C.  was partially supported by the National Science Center project DEC-2011/03/B/ST2/00136.

\section*{Supplementary material}

Let $\Lambda_t$ be an invertible map. We prove that $\Lambda_k$ is $k$-divisible if and only if the formula  (\ref{contr-k}) holds for all Hermitian $X \in M_k \ot \mathcal{B}(\mathcal{H}$).
We use the following \cite{Paulsen}
\begin{Lemma}
If $\Phi : \mathcal{B}(\mathcal{H}) \rightarrow \mathcal{B}(\mathcal{H})$ is trace-preserving, then $\Phi$ is positive if and only if
\begin{equation*}\label{}
  ||\Phi(X)||_1 \leq ||X||_1\ ,
\end{equation*}
for all Hermitian $X \in \mathcal{B}(\mathcal{H})$.
\end{Lemma}
Assuming that $\Lambda_t$ is $k$-divisible one has
\begin{eqnarray*}
 & &  \frac{d}{dt}\,||[\oper_k \ot \Lambda_t](X)||_1 \\
& & =  \lim_{\epsilon\rightarrow\, 0+} \frac{1}{\epsilon}\, \Big[ ||[\oper_k \ot \Lambda_{t+\epsilon}](X)||_1 - ||[\oper_k \ot \Lambda_t](X)||_1 \Big]  \\ & & =
  \lim_{\epsilon\rightarrow\, 0+} \frac{1}{\epsilon}\,\Big[ ||[\oper_k \ot V_{t+\epsilon,t}\, \Lambda_{t}](X)||_1 - ||[\oper_k \ot \Lambda_t](X)||_1 \Big]\\
   & & \leq  \lim_{\epsilon\rightarrow\, 0+} \frac{1}{\epsilon}\,\Big[ || [\oper_k \ot \Lambda_{t}](X)||_1 - ||[\oper_k \ot \Lambda_t](X)||_1 \Big] = 0 \ ,
\end{eqnarray*}
where we have used $k$-divisibility of $\Lambda_t$, i.e.
$$\Lambda_{t+\epsilon} = V_{t+\epsilon,t} \Lambda_t \ , $$
with $V_{t+\epsilon,t}$ being $k$-positive and the Lemma 1 for a  positive and trace-preserving map $\oper_k \ot V_{t,t+\epsilon}$
\begin{eqnarray*}
  ||[\oper_k \ot V_{t+\epsilon,t}\, \Lambda_{t}](X)||_1 &=& ||[\oper_k \ot V_{t+\epsilon,t}] [(\oper_k \ot \Lambda_{t})(X)] ||_1  \\
   &\leq& ||[\oper_k \ot \Lambda_{t}](X) ||_1 \ .
\end{eqnarray*}
To prove the converse result let $ Y = [\oper_k \ot \Lambda_{t}](X)$. Now, if $\frac{d}{dt}\,||[\oper_k \ot \Lambda_t](X)||_1 \leq 0$, then
$$ ||[\oper_k \ot V_{t+\epsilon,t}](Y)||_1  \leq ||Y||_1 \ .  $$
Assuming that $\Lambda_t$ is invertible we proved that $\oper_k \ot V_{t+\epsilon,t}$ is a contraction on all Hermitian elements $Y  \in M_k \ot \mathcal{B}(\mathcal{H})$ and hence, due to the Lemma 1, $\oper_k \ot V_{t+\epsilon,t}$ is positive or, equivalently, $V_{t+\epsilon,t}$ is $k$-positive.

\end{document}